\documentclass[aps,prl,twocolumn,amsmath,amssymb,superscriptaddress,floatfix]{revtex4-1}
\usepackage{amsfonts}
\usepackage[T1]{fontenc}
\usepackage{amsmath,amsbsy,amssymb,graphicx}
\usepackage{times}
\usepackage{amsmath}
\usepackage{amssymb}
\usepackage{graphicx}
\usepackage{color}
\usepackage{ulem}

\let\mathbf=\boldsymbol
\def\bA{\mathbf{A}}
\def\bk{\mathbf{k}}
\def\bq{\mathbf{q}}
\def\bg{\mathbf{g}}
\def\bB{\mathbf{B}}
\def\bE{\mathbf{E}}
\def\bv{\mathbf{v}}
\def\bj{\mathbf{j}}
\def\EF{E_\mathrm{F}}
\def\kF{k_\mathrm{F}}

\def\Ec{E_\mathrm{c}}
\def\kB{k_\mathrm{B}}
\def\Tc{T_\mathrm{c}}
\def\muB{\mu_\mathrm{B}}
\def\ER{E_\mathrm{R}}

\begin{document}

\title{Nonreciprocal current in noncentrosymmetric Rashba superconductors}

\author{Ryohei Wakatsuki}
\affiliation{Department of Applied Physics, University of Tokyo, Hongo 7-3-1, 113-8656, Japan}
\author{Naoto Nagaosa}
\affiliation{Department of Applied Physics, University of Tokyo, Hongo 7-3-1, 113-8656, Japan}
\affiliation{RIKEN Center for Emergent Matter Science (CEMS), Wako 351-0198, Japan}

\begin{abstract}
{\bf
Noncentrosymmetric superconductors with broken inversion symmetry $P$ offer
rich physical phenomena such as the upper critical magnetic field $H_{c2}$ beyond the 
Pauli limit and magnetoelectric effect. The relativistic spin-orbit interaction (SOI)
plays an essential role in these novel phenomena, which lifts the Kramers degeneracy
at each $\bk$-point, and leads to the mixing of the spin singlet-even parity 
and spin triplet odd-parity parings. On the other hand, 
the time-reversal symmetry $T$ relates the two states at $\bk$ and $-\bk$ points 
with the opposite spins, and the external magnetic field $B$ further breaking $T$
leads to the directional dependence of the nonlinear resistivity called magnetochiral anisotropy (MCA).
Here we demonstrate theoretically that the two-component nature of the order
parameter, i.e., even and odd pairings, leads to the gigantic enhancement of
    the MCA in the fluctuation region, in Rashba superconductor as a representative 
example of noncentrosymmetric system. This reveals the superconducting analogue
of "ferroelectric" appearing in the transport phenomena. 
}
\end{abstract}

\maketitle

The lack of spatial inversion $P$ and time-reversal $T$
symmetries is the fundamental issue in condensed matters. 
The breaking of the former allows the ferroelectric in insulator, while 
that of the latter leads to the magnetism. Multiferroic insulators with broken both
$P$ and $T$ is attracting intensive recent interests from the viewpoint
of various magnetoelectric effect and nonreciprocal effects \cite{Spaldin, Tokura}.
In metals with broken $P$, i.e., polar metals, one cannot define the 
electric polarization, and the effect of noncentrosymmetry is much less trivial.
However, the nonreciprocal responses in noncentrosymmetric metals and 
superconductors are the focus of recent studies \cite{Rikken1, Rikken2, Pop, Krstic, Rikken3, Morimoto, Ideue, Wakatsuki}.

In solids, the electronic states are described by Bloch wavefunctions whose 
energy eigenvalue is $\varepsilon_\sigma(\bk)$  with $\sigma$ being the 
spin component and $\bk$ the crystal momentum.  
The spatial inversion transforms $(\bk, \sigma)$ to $(-\bk, \sigma)$,
while the time-reversal $(\bk, \sigma)$ to $(-\bk, \bar{\sigma})$
($\bar{\sigma}$ is the opposite spin component to $\sigma$).
Therefore, when both $P$ and $T$ symmetries are broken, 
the electron pair with opposite momenta
is no longer related, and asymmetry between forward and backward appears.
In such situation, nonreciprocal charge current, whose conductivity depends
on the direction, can exist.
If we assume that the time-reversal symmetry breaking originates from 
the magnetic field $B$, the resistivity is traditionally expressed as
\begin{equation}
    R = R_0 \left(1+\gamma B I\right),
\end{equation}
where $I$ is the current, $B$ is the magnetic field, and $\gamma$ represents the 
nonreciprocity. This effect is named "magnetochiral anisotropy" (MCA).
There are several experiments on the MCA in normal state 
systems\cite{Rikken1, Rikken2, Pop, Krstic, Rikken3, Morimoto, Ideue}, and the typical value 
of $\gamma$ is $10^{-3} \sim 10^{-2}$ $\mathrm{T^{-1}A^{-1}}$.
The MCA has been studied also in superconducting fluctuation regime, 
where the thermal fluctuation of the superconducting order parameter creates charge 
current above the critical temperature \cite{Skocpol, Larkin}. In the monolayer transition 
metal dichalcogenides MoS$_2$\cite{Wakatsuki}, a quite large MCA 
$\gamma \sim 10^{3}$ $\mathrm{T^{-1}A^{-1}}$ has been observed, although 
the MCA in its normal state is almost zero. This drastic enhancement of the MCA stems from the energy scale difference between the Fermi energy $\EF$ 
and the superconducting gap $\Delta$. The theoretical analysis of the fluctuation 
of this material, however, is based on the warping of the Fermi surface, and does not
take into account the essential feature of the noncentrosymmetric superconductor, i.e.,
the mixing of the spin singlet-even parity and spin triplet-odd parity 
pairings\cite{BauerSigrist, Yip, Edelstein, Gorkov, Samokhin}, which will play the central role in the analysis below. 


In this paper, we study the nonreciprocal fluctuation current in Rashba superconductors. 
In order to treat the parity mixing appropriately, we employ the two-component 
Ginzburg--Landau (GL) theory.
We show the drastic enhancement of the MCA, which stems from the energy scale difference between $\EF$ and $\Delta$ similar to the case of MoS$_2$.
However, the two-component nature of the superconductivity is essential in the present case,
which is analogous to the ferroelectricity where the mixing of s- and p-orbitals 
produces the electric polarization.
We also show that the nonreciprocal current has a unique electric 
and magnetic fields angle dependence due to the symmetry constraints for the higher rank
response tensor.

We start with the Rashba Hamiltonian 
which is given by \cite{Ideue}
\begin{equation}
    H_{\bk} = \xi_{\bk} + \alpha \left( k_x \sigma_y - k_y \sigma_x \right) - \muB \bB \cdot \mathbf{\sigma}, \label{Eq:Rashba}
\end{equation}
where $\xi_{\bk} = \frac{\hbar^2 \bk^2}{2m} - \EF$ is the dispersion 
without the spin--orbit interaction with $\EF$ being the Fermi energy, $\alpha$ is the 
Rashba parameter, $\bB$ is the magnetic field, and $\mathbf{\sigma}$ 
are the Pauli matrices. We have assumed that the $g$-factor is 2.
Its eigenenergies are
\begin{equation}
\xi_{\pm\bk} = \xi_{\bk} \pm \sqrt{\left(\alpha k_y + B_x\right)^2
+\left(\alpha k_x - B_y\right)^2}. \label{Eq:dispersion}
\end{equation}
Now we mention the MCA in the normal state of Rashba system.
Due to the helical spin structure in the momentum space, the band is distorted along the direction perpendicular to the magnetic field.
Because of such asymmetry, when the electric field is applied 
perpendicular to the magnetic field, the nonreciprocal current occurs along 
the electric field direction.
According to Ref. \cite{Ideue}, the MCA exists if the Fermi energy is below the crossing point of the bands ($\EF < 0$).
The amplitude of the MCA is
\begin{equation}
    W \gamma_\mathrm{N} = \frac{3\pi\muB\hbar^2}{2e\sqrt{m}}\frac{\mathrm{sign}\left(\alpha\right)}{\ER\left(\ER-2\left|\EF\right|\right)^{3/2}}, \label{Eq:gammaN}
\end{equation}
with $W$ being the sample width and $\ER = \frac{m\alpha^2}{\hbar^2}$ being the energy splitting at the shifted momentum due to the Rashba spin--orbit interaction.

Now we consider the superconductivity of Rashba system \cite{BauerSigrist, Samokhin}.
For even parity attractive interaction, we assume the standard BCS type onsite attractive interaction,
\begin{equation}
    H_\mathrm{int} = -V^\mathrm{g} \sum_{\bk \bk'} c^\dagger_{\bk \uparrow}c^\dagger_{-\bk \downarrow} c_{-\bk' \downarrow}c_{\bk' \uparrow},
\end{equation}
with $c^\dagger_{\bk \sigma}$ and $c_{\bk \sigma}$ being the creation and 
annihilation operators of the electron with momentum $\bk$ and spin $\sigma$.
In general, the odd parity part is
\begin{equation}
    -\sum_{\bk\bk'}V_{ij}^\mathrm{u}\left(\bk, \bk'\right)\left(i\sigma_i\sigma_2\right)_{\alpha\beta}\left(i\sigma_j\sigma_2\right)_{\gamma\delta} c^\dagger_{\bk \alpha}c^\dagger_{-\bk \beta} c_{-\bk' \gamma}c_{\bk' \delta},
\end{equation}
with $V^\mathrm{u}_{ij}\left(\bk, \bk'\right)$ being an odd function with respect to $\bk$ and $\bk'$, and invariant under the crystal symmetry transformations. For simplicity, we assume the simplest case $V^\mathrm{u}_{ij}\left(\bk, \bk'\right) = V^\mathrm{u}\hat{\gamma}_i\left(\bk\right) \hat{\gamma}_j\left(\bk'\right)$ with $\hat{\gamma}\left(\bk\right) = \frac{1}{k}\left(-k_y, k_y\right)$ in the Rashba system.
Then, the interaction Hamiltonian in the band basis reads to
\begin{equation}
    H_\mathrm{int} = -\sum_{\bk \bk' \lambda \lambda'}
        t_{\bk \lambda} t_{\bk' \lambda'}^* \hat{g}_{\lambda \lambda'}
        \psi^\dagger_{\bk \lambda} \psi^\dagger_{-\bk \lambda}
        \psi_{-\bk' \lambda'} \psi_{\bk' \lambda'}, \label{Eq:Hint}
\end{equation}
where $\Psi^\dagger_{\bk \lambda}$ and $\Psi_{\bk \lambda}$ are the creation and annihilation operators with the band index $\lambda = \pm$, and $t_{\bk\lambda} = \lambda i \mathrm{e}^{i\phi_{\bk}}$ with $\phi_{\bk} = \mathrm{arg} \bk$.
The $\bk$-independent matrix $\hat{g}$ is
\begin{equation}
    \hat{g} = 
    \begin{pmatrix}
        g_1 & g_2 \\
        g_2 & g_1
    \end{pmatrix},
\end{equation}
with  $g_1 = \left(V^\mathrm{g}+V^\mathrm{u}\right)/4$ $(> 0)$ and $g_2 = \left(V^\mathrm{g}-V^\mathrm{u}\right)/4$.
In this paper, we focus on two regimes.
(1) $|V^\mathrm{u}| \ll |V^\mathrm{g}|$ or $g_2 \approx g_1$ case. If $g_2 = g_1$, the order parameter is purely singlet, and we consider the deviation from the limit by expanding with respect to the parameter $r_\mathrm{t} = \frac{g_1-g_2}{g_1}$, which is proportional to the triplet mixing.
(2) $|V^\mathrm{u}| \gg |V^\mathrm{g}|$ or $g_2 \approx -g_1$ case. If $g_2 = -g_1$, the order parameter is purely triplet, and we consider the deviation from the limit by expanding with respect to the parameter $r_\mathrm{s} = \frac{g_1+g_2}{g_1}$, which is proportional to the singlet mixing.

In order to calculate the superconducting fluctuation current slightly above the critical 
temperature, it is convenient to employ the GL theory.
The free energy quadratic with respect to the order parameters can be obtained by the equation \cite{BauerSigrist}
\begin{widetext}
\begin{equation}
    F = \int\frac{d^2\bq}{\left(2\pi\right)^2} \left[\sum_{\lambda \lambda'}
\Psi^*_{\lambda\bq} \left(\hat{g}^{-1}\right)_{\lambda \lambda'}\Psi_{\lambda' \bq}
        -\sum_{\lambda}T\sum_{\omega_n}\int\frac{d^2\bk}{\left(2\pi\right)^2}
G_\lambda\left(\bk, i\omega_n\right)
        G_\lambda\left(-\bk+\bq, -i\omega_n\right)|\Psi_{\lambda\bq}|^2
        \right],
\end{equation}
\end{widetext}
where $\Psi_{\lambda\bq}$ is the order parameter and $G_\lambda\left(\bk, i\omega_n\right) 
= \left(i\omega_n - \xi_{\lambda\bk}\right)^{-1}$ is the non-interacting normal Green's function.
We set the Boltzmann constant $\kB=1$.

Firstly, we assume $\EF > 0$, and we will soon show that nonreciprocal current vanishes for $\EF < 0$ in Eq. (\ref{Eq:j2}) below.
After some calculations (see Supplementary Information), we obtain
\begin{equation}
    F = \int\frac{d^2\bk}{\left(2\pi\right)^2}\sum_{\lambda \lambda'} \Psi^*_\lambda 
    \left[ \left(\hat{g}^{-1}\right)_{\lambda \lambda'} + \delta_{\lambda\lambda'}N_{\lambda} 
\left(S_1 - L_{\lambda\bk}\right) \right] \Psi_{\lambda'}, \label{Eq:free}
\end{equation}
\begin{eqnarray}
    L_{\lambda\bk} &=& K_\lambda \bk^2 - \lambda R_\lambda \left(B_y k_x - B_x k_y\right), \\
    S_1 &=& \log \frac{2\mathrm{e}^{\gamma_\mathrm{E}} \Ec}{\pi T},
\end{eqnarray}
with $\delta_{\lambda \lambda'}$, $\gamma_\mathrm{E}$, and $\Ec$ being the Kronecker delta, Euler constant, and cutoff energy respectively. The density of states $N_{\lambda}$ and the other coefficients $K_{\lambda}$ and $R_{\lambda}$ are given in Supplementary Information.
The critical temperatures are obtained by solving
\begin{equation}
    \det\left(\hat{g}^{-1} - \hat{N} S_1\left(\Tc\right)\right) = 0,
\end{equation}
with $\hat{N}_{\lambda\lambda'}=\delta_{\lambda\lambda'}N_\lambda$.
It results in
\begin{equation}
    \frac{1}{S_1\left(\Tc\right)} = \frac{g_1\left(N_-+N_+\right)}{2}
    \pm \sqrt{\left(\frac{g_1\left(N_--N_+\right)}{2}\right)^2+g_2^2N_-N_+}. \label{Eq:char}
\end{equation}
Due to the form of the interaction ($g_1 \approx g_2$ for the singlet dominant case and $g_1 \approx -g_2$ for the triplet dominant case), the solution with the plus sign has much higher critical temperature. 
Hence, we can ignore the order parameter with the lower critical temperature when we calculate 
the fluctuation current.

The fluctuation current can be obtained by evaluating the equation \cite{Schmid, Wakatsuki}
\begin{widetext}
\begin{equation}
    \bj = -T \sum_{\bk} C\left.\frac{\partial \eta\left(\bk + 2 e \bA\right)}{\partial \bA}\right|_{\bA=\mathbf{0}}\int_{-\infty}^0 du \exp\left[-C\int_u^0dt \eta\left(\bk - 2e \bE t\right) \right], \label{Eq:formula}
\end{equation}
\end{widetext}
where $\eta$ is the eigenvalue of the matrix in Eq. (\ref{Eq:free}) with the higher critical temperature, and $C = \frac{32\Tc}{\pi\hbar\left(N_-+N_+\right)} + O\left(r_\mathrm{t, s}\right)$. It is noted that the factor $C$ should contain a $r_{\mathrm{t,s}}$-dependent correction from the relaxation time of order parameters in the time-dependent GL theory. However, we ignore it because it does not affect the $\gamma$ value in the lowest order of $r_{\mathrm{t,s}}$.
As in the case of the normal state, we assume that the electric and magnetic fields are applied along the $x$ and $y$ directions respectively, and evaluate the current along the $x$ direction up to $O\left(B_y E_x^2\right)$.
We will discuss the case of general fields configurations later.
After the integration in Eq. (\ref{Eq:formula}) is carried out (we employed Mathematica), the relation Eq. (\ref{Eq:char}) is used to simplify the equation. The result is
\begin{widetext}
\begin{eqnarray}
    j_x &=& \sigma^{\left(1\right)}E_x + \sigma^{\left(2\right)} E_x^2, \label{Eq:j} \\
    \sigma^{\left(1\right)} &=& \frac{e^2}{16\hbar \varepsilon}, \label{Eq:j1}\\
    \sigma^{\left(2\right)} &=& \frac{\pi e^3 B_y r_{\mathrm{t,s}}}{128\hbar\varepsilon^2}
    \frac{N_-N_+\left(K_-N_--K_+N_+\right)\left(K_-R_++K_+R_-\right)}{S_1\left(\Tc\right)\Tc\left(N_-+N_+\right)\left(K_-N_-+K_+N_+\right)^2}, \label{Eq:j2}
\end{eqnarray}
\end{widetext}
in the lowest order of $r_{\mathrm{t,s}}$.
Here, we have defined the reduced temperature $\varepsilon = \frac{T-\Tc}{\Tc}$.
The linear coefficient $\sigma^{\left(1\right)}$ is the conventional form of the fluctuation conductivity in two-dimensional superconductors. The nonlinear coefficient $\sigma^{\left(2\right)}$ grows faster than $\sigma^{\left(1\right)}$ toward the critical temperature as in the case of MoS$_2$ \cite{Wakatsuki}.
It is noted that the parity mixing is essential for the nonreciprocal current, which vanishes for $r_{\mathrm{t,s}}=0$.

We mention the case when the Fermi energy is below the crossing point of the bands ($\EF < 0$). In this case, because the density of states from the upper band is zero, the nonreciprocal current vanishes, whereas, the normal current contribution exists \cite{Ideue}.

For $\EF > 0$, the $\gamma$ value expressed with the microscopic parameters is
\begin{equation}
    W \gamma_\mathrm{S} = 
    \frac{\sigma^{(2)}}{B_y \left(\sigma^{(1)}\right)^2} = 
        \frac{\pi\hbar^2\muB}{e\sqrt{m}}
        \frac{r S_3 \EF \sqrt{\ER}\mathrm{sign}\left(\alpha\right)}{S_1\Tc\left(2\EF+\ER\right)}, \label{Eq:gammaS}
\end{equation}
with $S_3 = \frac{7\zeta\left(3\right)}{4\pi^2\Tc^2}$. We have used the relation between $\sigma^{(1)}$, $\sigma^{(2)}$, and $\gamma$ as shown in Ref. \cite{Wakatsuki}.
We compare the $\gamma$ values in the normal regime (Eq. (\ref{Eq:gammaN})) and the superconducting fluctuating regime (Eq. (\ref{Eq:gammaS})).
In the normal regime, the nonreciprocal current exists in the case of $\EF < 0$. We assume that the strength of the spin--orbit interaction is comparable with the Fermi energy ($\ER \approx \left|\EF\right|$) because it is difficult to realize $\EF < 0$ with a small $\ER$. Then, we obtain
\begin{equation}
    W \gamma_\mathrm{N} \sim \frac{\muB \hbar^2}{e\sqrt{m}}\frac{1}{\left|\EF\right|^{5/2}}. \label{Eq:gamNap}
\end{equation}

    In the superconducting fluctuation regime, the nonreciprocal fluctuation current exists in the case of $\EF > 0$, which is opposite to the normal state. With the same assumption for the normal state, we obtain
\begin{equation}
    W \gamma_\mathrm{S} \sim \frac{\muB \hbar^2}{e\sqrt{m}}\frac{r \EF^{1/2}}{S_1 \Tc^3}. \label{Eq:gamSap}
\end{equation}

    From Eqs. (\ref{Eq:gamNap}) and (\ref{Eq:gamSap}), we conclude that the MCA is drastically enhanced in the superconducting fluctuation regime because of the huge energy scale difference between the Fermi energy $\EF$ and the critical temperature $\Tc$. This is similar to the proceeding results for MoS$_2$ \cite{Wakatsuki}.


\begin{figure}[t]
    \centerline{\includegraphics[width=0.5\textwidth]{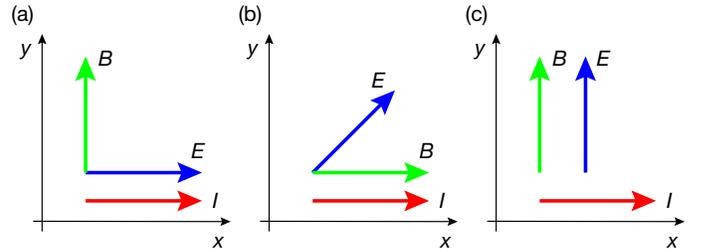}}
    \caption{
        (color online).
        The three fields configurations which correspond to (a) $\sigma_{xyxx}$, (b) $\sigma_{xxxy}$, and (c) $\sigma_{xyyy}$. $B$, $E$, and $I$ in the figures represent the electric field, magnetic field, and nonreciprocal current respectively.
    }
\label{Fig:direction}
\end{figure}

\begin{figure}[t]
    \centerline{\includegraphics[width=0.4\textwidth]{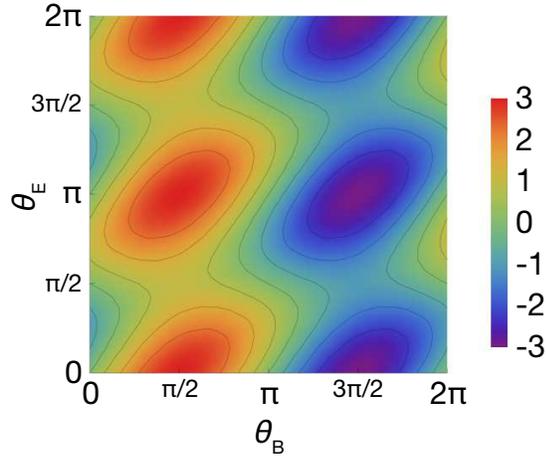}}
    \caption{
        (color online).
        The electric and magnetic fields angle dependence of $j^{\left(2\right)}$.
        $\theta_\mathrm{B}$ ($\theta_\mathrm{E}$) represents the angle between the magnetic (electric) field and the nonreciprocal current.
        The amplitude is normalized by $\sigma_{xyyy}BE^2$.
    }
\label{Fig:angle}
\end{figure}

We finally mention the electric and magnetic fields angle dependence of the nonreciprocal current. If we express the second order current as $j_i = \sigma_{ijkl} B_j E_k E_l$, the coefficient $\sigma_{ijkl}$ is the pseudo tensor consistent with the crystal symmetry. Our model Eq. (\ref{Eq:Rashba}) possesses $C_\infty$ symmetry and arbitrary in-plane mirror symmetries, which impose the restrictions that among $\sigma_{xjkl}$, only $\sigma_{xxxy}\left(=\sigma_{xxyx}\right)$, $\sigma_{xyxx}$, and $\sigma_{xyyy}$ can be finite (corresponding configurations are shown in Fig. \ref{Fig:direction}), and $\sigma_{xyyy}=2\sigma_{xxxy}+\sigma_{xyxx}$ and $\sigma_{yjkl}=-\sigma_{xjkl}$ are satisfied.
According to calculations the same as that for $\sigma_{xyxx}$ above, we obtain $\sigma_{xxxy} = -\frac{1}{3}\sigma_{xyxx}$ and $\sigma_{xyyy} = \frac{1}{3}\sigma_{xyxx}$, which satisfy the above conditions.
If we define the angle between the current and magnetic (electric) field as $\theta_\mathrm{B} (\theta_\mathrm{E})$, the nonreciprocal current is
\begin{equation}
    j^{(2)} = \sigma_{xyyy}\left(2\sin\theta_\mathrm{B}+\sin\left(\theta_\mathrm{B}-2\theta_\mathrm{E}\right)\right) B E^2, \label{Eq:jangle}
\end{equation}
whose dependence in the $\left(\theta_\mathrm{B}, \theta_\mathrm{E}\right)$ plane is shown in Fig. \ref{Fig:angle}.
It is noted that the normal state has the same angle dependence although it has not been discussed in the previous paper \cite{Ideue}.
Realistic materials do not have such high symmetries, however, the above discussion should be applicable if the Fermi surface is almost circular.

We have investigated the MCA of the Rashba system in the superconducting 
fluctuation regime.
The main result is the explicit form of the $\gamma$ value shown in Eq. (\ref{Eq:gammaS}).
Now, we estimate the value of $\gamma$ for BiTeBr, whose MCA in its normal state
has been measured \cite{Ideue}.
Because the superconductivity in BiTeBr has not been observed, we assume that the superconductivity is induced by the superconducting proximity effect.
In BiTeBr, the effective mass is $m=0.15m_{\mathrm{e}}$ \cite{Lee}, the Rashba parameter is $\alpha=2.00\mathrm{eV}${\AA} \cite{Sakano}, and the $g$-factor is $g=60$ \cite{Park}.
In the normal state with $\EF = -0.01 \mathrm{eV}$, the amplitude of the MCA is $\gamma_\mathrm{N} \approx 2 \times 10^{-5} \mathrm{T^{-1}A^{-1}m}$.
In the superconducting fluctuation regime, by assuming $\EF = 0.01 \mathrm{eV}$, $\Tc=1\mathrm{K}$, $\Ec=400\mathrm{K}$, and $r_\mathrm{t} = 0.1$, we obtain $\gamma_\mathrm{S} \approx 8 \times 10^{-2} \mathrm{T^{-1}A^{-1}m}$.
This value is much larger than that of the normal state, and if we assume a realistic sample width $W = 1\mu\mathrm{m}$, we have $\gamma_\mathrm{S} \approx 10^5 \mathrm{T^{-1}A^{-1}}$, which is quite large compared with the preceding results.

Such a huge enhancement of the MCA originates from the energy scale difference between 
the Fermi energy $\EF$ and the critical temperature $\Tc$ as indicated in 
Eqs. (\ref{Eq:gamNap}) and (\ref{Eq:gamSap}). This phenomenon is similar to the case of superconducting MoS$_2$ 
\cite{Wakatsuki}, in which the large MCA stems from the trigonal warping term due to 
its three-fold rotational symmetry. However, the MCA
originates from the parity mixing of the order parameter in the present case.

We have also shown the unique fields angle dependence of the nonreciprocal current, which is summarized in Fig. \ref{Fig:direction}.
It originates from the symmetry constraints of the higher rank response tensor.
Especially, if the Fermi surface is almost circular and  well approximated by our model, the fields angle dependence is given in Eq. (\ref{Eq:jangle}) and shown in Fig. \ref{Fig:angle}.
As candidate materials with a circular Fermi surface and giant Rashba splitting, we propose the BiTeX (X=I, Br, Cl) \cite{Ishizaka, Sakano}. We also expect that the interface of Au(111) \cite{LaShell} or Bi/Ag(111) alloy \cite{Ast} work well. However, the superconductivity should be induced by the proximity effect because they are not superconducting.

Experimentally, the nonreciprocal current can be observed simply by measuring second order harmonic voltage drop under a fixed a.c. current.
With such a simple method, we can observe the nontrivial second order response which reflects the crystal symmetry or the Hall response of the nonlinear current shown in Fig. \ref{Fig:direction}(c). It is also possible to determine the sign of $\alpha$ from the sign of the $\gamma$ value.
Moreover, we may estimate the amplitude of $r_\mathrm{t,s}$, which is the ratio between the even and odd parity attractive interactions by using the measured $\gamma_\mathrm{S}$ value.

\section*{Acknowledgements}
The authors thank Y. Saito, T. Ideue, and Y. Iwasa for valuable discussions.
R.W. was supported by the Grants-in-Aid for Japan Society for the Promotion of Science No. JP15J09045.
N.N. was supported by Ministry of Education, Culture, Sports, Science, and Technology Nos. JP24224009 and JP26103006, the Impulsing Paradigm Change through Disruptive Technologies Program of Council for Science, Technology and Innovation (Cabinet Office, Government of Japan), and Core Research for Evolutionary Science and Technology (CREST) No. JPMJCR16F1.

\begin{widetext}
\appendix
\section{Microscopic derivation of the Ginzburg--Landau theory}
Following Ref. \cite{BauerSigrist}, we review the derivation of the GL free energy of the Rashba superconductor from its microscopic Hamiltonian.
Firstly, we assume $\EF > 0$. What we have to calculate is
\begin{equation}
    F = \int\frac{d^2\bq}{\left(2\pi\right)^2} \left[\sum_{\lambda \lambda'}\Psi^*_{\lambda\bq} \left(\hat{g}^{-1}\right)_{\lambda \lambda'}\Psi_{\lambda' \bq}
        -\sum_{\lambda}T\sum_{\omega_n}\int\frac{d^2\bk}{\left(2\pi\right)^2}G_\lambda\left(\bk, i\omega_n\right)
        G_\lambda\left(-\bk+\bq, -i\omega_n\right)|\Psi_{\lambda\bq}|^2
        \right].
\end{equation}
We consider the Rashba system with in-plane magnetic field. The Hamiltonian is Eq. (2) in the main text. The energy dispersion (Eq. (3) in the main text) and the Green's function can be approximated with the small Zeeman energy as following equations.
\begin{eqnarray}
    \xi_\lambda\left(\bk, \bB\right) &=& \xi\left(\bk\right) + \lambda|\bg\left(\bk\right)-\muB\bB|
    \approx \xi_\lambda\left(\bk\right) - \lambda \muB \hat{\bg}\left(\bk\right)\cdot \bB, \\
    G_\lambda\left(k\right) &=& \frac{1}{i\omega_n-\xi_\lambda\left(\bk, \bB\right)}
    \approx \frac{1}{i\omega_n - \xi_\lambda\left(\bk\right) + \lambda \muB \hat{\bg}\left(\bk\right) \cdot \bB},
\end{eqnarray}
with $\xi_\lambda\left(\bk\right) = \xi\left(\bk\right) + \lambda \left|\bg\left(\bk\right)\right|$ and $\bg\left(\bk\right) = \alpha \left(-\frac{k_y}{k}, \frac{k_x}{k}\right)$, and we assume $\alpha > 0$ for simplicity.
Therefore,
\begin{eqnarray}
    && \int \frac{d^2\bk}{\left(2\pi\right)^2} G_\lambda\left(\bk, i\omega_n\right) G_\lambda\left(-\bk+\bq, -i\omega_n\right) \\
    &=& \int \frac{d^2\bk}{\left(2\pi\right)^2} \frac{1}{i\omega_n-\xi_\lambda\left(\bk\right)+\lambda\muB\hat{\bg}\left(\bk\right)\cdot\bB} \frac{1}{-i\omega_n-\xi_\lambda\left(\bk-\bq\right)-\lambda\muB\hat{\bg}\left(\bk-\bq\right)\cdot\bB} \\
    &\approx& \int \frac{d^2\bk}{\left(2\pi\right)^2} \frac{1}{i\omega_n-\xi_\lambda\left(\bk\right)+\lambda\muB\hat{\bg}\left(\bk\right)\cdot\bB} \frac{1}{-i\omega_n-\xi_\lambda\left(\bk\right)+\hbar\bq\cdot\bv_\lambda\left(\bk\right)-\lambda\muB\hat{\bg}\left(\bk-\bq\right)\cdot\bB} \\
    &\approx& N_\lambda \int d\xi \left< \frac{1}{i\omega_n-\xi+\lambda\muB\hat{\bg}\left(\bk\right)\cdot\bB} \frac{1}{-i\omega_n-\xi+\hbar\bq\cdot\bv_\lambda\left(\bk\right)-\lambda\muB\hat{\bg}\left(\bk-\bq\right)\cdot\bB} \right>_{\lambda} \\
    &=& \pi N_\lambda \left< \frac{1}{|\omega_n| + i \Omega_\lambda\left(\bk, \bq\right) \mathrm{sign}\left(\omega_n\right)} \right>_{\lambda},
\end{eqnarray}
where $\left< \cdot \right>_{\lambda}$ represents the average over the Fermi surface with the band index $\lambda$, and
\begin{equation}
    N_\lambda = \frac{m}{2\pi\hbar^2}\left(1-\lambda\frac{\sqrt{\ER}}{\sqrt{2\EF+\ER}}\right)
\end{equation}
is the density of states, and we have defined
\begin{equation}
    \Omega_\lambda\left(\bk, \bq\right) = \frac{1}{2}\hbar \bq\cdot\bv_\lambda\left(\bk\right)-\frac{\lambda\muB}{2}\left(\hat{\bg}\left(\bk\right)+\hat{\bg}\left(\bk-\bq\right)\right) \cdot \bB.
\end{equation}
Therefore,
\begin{eqnarray}
    T\sum_{\omega_n}\int\frac{d^2\bk}{\left(2\pi\right)^2}G_\lambda\left(\bk, i\omega_n\right)
        G_\lambda\left(-\bk+\bq, -i\omega_n\right)
        &\approx&\pi T N_\lambda \sum_{\omega_n} \left< \frac{1}{|\omega_n|} - \frac{\left(\Omega_\lambda\left(\bk, \bq\right)\right)^2}{|\omega_n|^3} \right>_{\lambda} \notag \\
        &=& N_\lambda \left[ S_1\left(T\right) - S_3\left(T\right) \left< \left(\Omega_\lambda\left(\bk, \bq\right)\right)^2 \right>_{\lambda} \right],
\end{eqnarray}
where
\begin{eqnarray}
    S_1\left(T\right) &=& \pi T \sum_{\omega_n} \frac{1}{|\omega_n|} = \log \frac{2\mathrm{e}^{\gamma_\mathrm{E}} \Ec}{\pi T}, \\
    S_3\left(T\right) &=& \pi T \sum_{\omega_n} \frac{1}{|\omega_n|^3} = \frac{7\zeta\left(3\right)}{4\pi^2T^2}.
\end{eqnarray}
Here, $\gamma_\mathrm{E}$ is the Euler constant and $\Ec$ is the energy cutoff which correspond to the Debye frequency.
Moreover,
\begin{eqnarray}
\left< \left(\Omega_\lambda\left(\bk, \bq\right)\right)^2 \right>_{\lambda}
    &=& \frac{1}{4} \left<\left(\hbar\bq \cdot \bv_\lambda\left(\bk\right) \right)^2\right>
    -\lambda \muB \left< \left(\hbar\bq \cdot \bv_\lambda\left(\bk\right)\right) \left(\hat{\bg}\left(\bk\right) \cdot \bB \right)\right>_{\lambda} \notag \\
    &=& \frac{1}{8}\left(\frac{\hbar^2k_{\mathrm{F}\lambda}}{m}+\lambda\alpha\right)^2 \bq^2
    - \frac{1}{2}\lambda \muB\left(\frac{\hbar^2k_{\mathrm{F}\lambda}}{m}+\lambda\alpha\right)\left(B_y q_x - B_x q_y\right),
\end{eqnarray}
where $k_{\mathrm{F}\lambda} = -\lambda m\alpha + \sqrt{\left(m\alpha\right)^2+2m\EF}$.
Finally, we obtain
\begin{eqnarray}
    F &=& \int \frac{d^2\bq}{\left(2\pi\right)^2}
        \sum_{\lambda \lambda'} \Psi^*_\lambda\left(\bq\right) \left(g^{-1}\right)_{\lambda \lambda'}\Psi_{\lambda'}\left(\bq\right) \notag \\
      &&  - \sum_\lambda N_\lambda \left[S_1 - \frac{1}{8}S_3\frac{1}{\kF^2}\left(4\EF^2+\ER^2\right)\bq^2 + \frac{1}{2}\lambda\muB S_3\frac{1}{\kF}\sqrt{4\EF^2+\ER^2}\left(B_yq_x-B_xq_y\right)\right]|\Psi_\lambda\left(\bq\right)|^2.
\end{eqnarray}
If we define
\begin{eqnarray}
    L_{\lambda\bk} &=& K_\lambda \bk^2 - \lambda R_\lambda \left(B_y k_x - B_x k_y\right), \\
    K_- &=& K_+ = \frac{\hbar^2 S_3 \left(2\EF+\ER\right)}{8m}, \\
    R_- &=& R_+ = \frac{\hbar \muB S_3 \sqrt{2\EF+\ER}}{2\sqrt{m}},
\end{eqnarray}
we have the compact form of the free energy,
\begin{equation}
    F = \int\frac{d^2\bk}{\left(2\pi\right)^2}\sum_{\lambda \lambda'} \Psi^*_\lambda 
    \left[ \left(\hat{g}^{-1}\right)_{\lambda \lambda'} + \delta_{\lambda\lambda'}N_{\lambda} 
\left(S_1\left(T\right) - L_{\lambda\bk}\right) \right] \Psi_{\lambda'},
\end{equation}
which is Eq. (10) in the main text.

The case of $\EF < 0$ can be calculated in a similar manner. Because the Fermi energy does not cross the upper band, only the $\left(--\right)$ component is finite in the $g$-independent part.
\begin{equation}
    F = \int\frac{d^2\bk}{\left(2\pi\right)^2}
    \left[
    \sum_{\lambda \lambda'} \Psi^*_\lambda \left(\hat{g}^{-1}\right)_{\lambda \lambda'} \Psi_{\lambda'}
    + \Psi^*_{-} \left[\left(N_1+N_2\right) \left(S_1\left(T\right) - K' \bk^2\right) + \left(N_1-N_2\right) R' \left(B_y k_x - B_x k_y\right)
    \right] \Psi_{-}
    \right],
\end{equation}
with
\begin{eqnarray}
    N_1 &=& \frac{m}{2\pi\hbar^2}\left(1+\frac{\sqrt{\ER}}{\sqrt{\ER-2\left|\EF\right|}}\right), \\
    N_2 &=& -\frac{m}{2\pi\hbar^2}\left(1-\frac{\sqrt{\ER}}{\sqrt{\ER-2\left|\EF\right|}}\right), \\
    K' &=& \frac{\hbar^2 S_3 \left(\ER-2\left|\EF\right|\right)}{8m}, \\
    R' &=& \frac{\hbar \muB S_3 \sqrt{\ER-2\left|\EF\right|}}{2\sqrt{m}}.
\end{eqnarray}
We note that for $\alpha < 0$, the sign of the term proportional to $\bB$ is inverted.

\end{widetext}

\end{document}